\documentclass[aps,preprint,amsmath,amssymb,showpacs]{revtex4}
\usepackage{graphicx}
\def\be{\begin{eqnarray}}
\def\en{\end{eqnarray}}
\def\non{\nonumber}
\def\la{\langle}
\def\ra{\rangle}

\def\prl{{ Phys. Rev. Lett.}~}

\def\bi{\bibitem}
\def\t{\perp}
\begin{document}

\title{\Large \bf Study of quark distribution amplitudes of $1S$ and $2S$ heavy quarkonium
states
 }

\author{ \bf  Chien-Wen Hwang\footnote{
t2732@nknucc.nknu.edu.tw}}

\affiliation{\centerline{Department of Physics, National Kaohsiung Normal University,} \\
\centerline{Kaohsiung, Taiwan 824, Republic of China}
 }


\begin{abstract}
In this paper, the quark distribution amplitudes of $1S$ and $2S$
heavy quarkonium states are studied in terms of Gaussian-type wave
functions. The transverse momenta $p_\perp$ integrals of the
formulae for the decay constant are performed analytically. Then the
quark distribution amplitudes are obtained. In addition, the
$\xi$-moments are also calculated. After fixing the relevant
parameters appearing in the quark distribution amplitude, the curves
of the quark distribution amplitude for $1S$ and $2S$ heavy
quarkonium states are plotted. Finally, the numerical results of
this approach are compared with the other theoretical predictions.
\end{abstract}
\maketitle %

\section{Introduction}
Since the discoveries of $J/\psi$ and $\Upsilon$ more than thirty
years ago, a great deal of information on heavy quarkonium levels
and their transitions has been accumulated \cite{PDG08}. On one
hand, the known levels of charmonium and bottomonium include $1S$,
$1P$, $1D$, $2S$, $2P$, $2D$, $3S$, and $4S$ states, with the labels
$S$, $P$, $D$ corresponding to relative orbital angular momentum $L$
= 0, 1, 2 between quark and antiquark. On the other hand, the
numerous transitions between heavy quarkonium states are classified
into strong and radiative decays, which shed light on aspects of
quantum chromodynamics (QCD) in both the perturbative and the
non-perturbative regimes (for a recent review see \cite{EGMR}).
Recently the more precise $\psi'$, $\eta_c'$, and $\eta_b$ mass
measurements have been reported \cite{KEDR,CLEO,BABAR,BABAR1}, and
the errors of their relevant decay widths have decreased
\cite{PDG08}. A thorough understanding of their properties, such as
their quark distribution amplitudes, which are the universal
non-perturbative objects, will be of great benefit when analyzing
the hard exclusive processes with heavy quarkonium production.

It has been known that heavy quarkonium is relevant for a
non-relativistic treatment \cite{QR}. Although non-relativistic QCD
(NRQCD) is a powerful theoretical tool for separating the high
energy modes from low energy contributions, in most cases the
calculation of low energy hadronic matrix elements has relied on
model-dependent non-perturbative methods. Therefore, various methods
have been employed in heavy quarkonium physics, such as lattice QCD,
quark-potential model, etc. \cite{EGMR}. The light-front quark model
offers many insights into the internal structures of the bound
states. In this study, heavy quarkonium is explored within a quark
model on the light front. Light-front QCD is a promising analytic
method for solving the non-perturbative problems of hadron physics
\cite{BPP}, and may be the only possible method by which the low
energy quark model and the high energy parton model can be
reconciled. For hard processes with a large momentum transfer,
light-front QCD reduces to perturbative QCD (pQCD) which factorizes
the physical quantity into a convolution of the hard scattering
kernel and the quark distribution amplitudes.

The basic ingredient in light-front QCD is the relativistic hadron
wave function. It generalizes the distribution amplitudes by
including the transverse momentum distributions, and it contains all
the information of a hadron from its constituents. The hadronic
quantities are represented by the overlap of wave functions and can
be derived in principle. The light-front wave function is manifestly
a Lorentz invariant as it is expressed in terms of the internal
momentum fraction variables which are independent of the total
hadron momentum. Moreover, the fully relativistic treatment of quark
spins and the center-of-mass motion can be carried out using the
so-called Melosh rotation \cite{LFQM,LFQMa,LFQMb}. This treatment
has been successfully applied to calculate many phenomenologically
important meson decay constants and hadronic form factors
\cite{Jaus1,Jausa, CCH1, Jaus2, CCH2, Hwang}. In addition, the
covariant light-front approach \cite{Jaus2, CCH2} has also been
applied to ground-state $s$-wave mesons, which include $1^1S_0$
pseudoscalar mesons $\eta_c, \eta_b$ and $1^3S_1$ vector mesons
$J/\psi, \Upsilon$ (see Ref. \cite{hwangwei}).

As mentioned above, the quark distribution amplitude or light-cone
wave function (LCWF) absorbs the non-perturbative dynamics and is
the key ingredient of any hard exclusive process with hadron
production. In the literature, there are many theoretical studies
\cite{Braguta1,Braguta2,Braguta3,BT,EGKLY,BC,CJ,BF} of this issue.
The main purpose of this study is the calculation of the quark
distribution amplitudes of pseudoscalar and vector heavy quarkonium
states by integrating the transverse momenta of momentum
distribution amplitudes within the light-front approach. As to the
so-called $\xi$-moments which parameterize the quark distribution
amplitudes, they are also calculated analytically.

The remainder of this paper is organized as follows. Section II
comprises brief reviews of the light-front framework and the
light-front analysis for the decay constants of pseudoscalar ($P$)
and vector ($V$) mesons; the processes $V\to P\gamma$ are given. In
Section III, the quark distribution amplitudes and the $\xi$-moments
are calculated. In Section IV, the numerical results and discussions
are presented. Finally, the conclusions are given in Section V.

\section{Formalism of covariant light-front approach}
In heavy quarkonium, the valence quarks have equal masses,
$m_1=m_2=m$, with $m$ the mass of heavy quark $c$ or $b$. Thus, the
formulae in this section lead to simplifications for the quarkonium
system.

The momentum of a particle is given in terms of the light-front
component by  $k=(k^-, k^+, k_\t)$ where $k^{\pm}=k^0\pm k^3$ and
$k_\t=(k^1, k^2)$, and the light-front vector is written as $\tilde
k=(k^+, k_{\t})$. The longitudinal component $k^+$ is restricted to
positive values, i.e., $k^+> 0$ for the massive particle. In this
way, the physical vacuum of light-front QCD is trivial except for
the zero longitudinal momentum modes (zero-mode). A meson with total
momentum $P$ and two constituents, quark and anti-quark whose
momenta are $p_1$ and $p_2$, respectively will be studied. In order
to describe the internal motion of the constituents, it is crucial
to introduce the intrinsic variables $(x, p_{\t})$ through
 \be
 && p_1^+=x P^+,  ~~~~~~~~~~~  p_{1\t}=x P_{\bot}+ p_{\t};  \non \\
 && p_2^+=(1-x) P^+,   ~~~  p_{2\t}=(1-x) P_{\bot}- p_{\t},
 \en
where $x$ is the light-front momentum fraction. The invariant mass
$M_0$ of the constituents and the relative momentum in $z$ direction
$p_z$ can be written as
 \be \label{eq:Mpz}
  M_0^2=\frac{p_{\t}^2+m^2}{x(1-x)}, \qquad ~~~
  p_z=\left(x-\frac{1}{2}\right)M_0.
 \en
The invariant mass $M_0$ of $q\bar q$ is generally different from
the mass $M$ of meson which satisfies $M^2=P^2$. This is due to the
fact that the meson, quark and anti-quark cannot be on-shell
simultaneously. The momenta $p_{\t}$ and $p_z$ constitute a momentum
vector $\vec p=(p_{\t}, p_z)$, which represents the relative momenta
in the transverse and $z$ directions, respectively.  The energy of
the quark and antiquark $e_1=e_2\equiv e$ can be obtained from their
relative momenta,
 \be
 e=\sqrt{m^2+p_{\t}^2+p_z^2}.
 \en
It is straightforward to find that
 \be
 x=\frac{e-p_z}{2e},~ e=\frac{M_0}{2}.
 \en

As shown in Ref.~\cite{CM69}, one can pass to the light-front
approach by integrating the $p^-$ component of the internal momentum
in the covariant Feynman momentum loop integrals. The Feynman rules
for the meson-quark-anti-quark vertices were then needed to
calculate the amplitudes which related to the decay constant and M1
transition. In the following formulations, we follow the notation in
\cite{CCH2}. The vertices $\Gamma_M$ for the incoming meson $M$ are
given as
 \be \label{eq:HM}
 H_P \gamma_5  \qquad \qquad   \qquad \qquad
 &&{\rm for~ }P, \non \\
 iH_V\Big [ \gamma_{\mu}-\frac{1}{W_V}(p_1-p_2)_{\mu} \Big ]
  \qquad &&{\rm for~ }V.
 \en
The Feynman rules which are derived from quantum field theory and
Eq. (\ref{eq:HM}) can be used to write down the relevant matrix
elements. The integration of the $p^-$ component will force the
quark or anti-quark to be on its mass shell. The specific form of
the covariant vertex function for the on-shell (anti)quark can be
determined by comparing it to the conventional vertex function,
which can be written as
 \be
 |M(P, ^{2S+1}L_J, J_z)\rangle
 =\int &&\{d^3p_1\}\{d^3p_2\} ~2(2\pi)^3 \delta^3(\tilde
 P -p_1-\tilde p_2)~\nonumber\\
 &&\times \sum_{\lambda_1,\lambda_2}
 \Psi^{SS_z}(\tilde p_1,\tilde p_2,\lambda_1,\lambda_2)~
 |q_1(p_1,\lambda_1) \bar q_2(p_2,\lambda_2)\rangle.
 \label{lfmbs}
 \en
The momentum-space wave function $\Psi^{SS_z}$ can be expressed as
 \be
 \Psi^{SS_z}(\tilde p_1,\tilde
 p_2,\lambda_1,\lambda_2)
                = R^{SS_z}_{\lambda_1\lambda_2}(x,p_\bot)~ \phi_(x, p_\bot),
 \en
where $\phi(x,p_\perp)$ is the light-front momentum distribution
amplitude for the $s$-wave meson and can be chosen to be
normalizable, i.e., it satisfies
 \be \label{eq:Norm1}
 \int \frac{dx d^2 p_{\bot}}{2 (2\pi)^3}
   |\phi(x,p_{\bot})|^2=1,
 \en
and $R^{SS_z}_{\lambda_1\lambda_2}$ constructs a state of definite
spin ($S,S_z$) out of light-front helicity ($\lambda_1,\lambda_2$)
eigenstates. In practice, it is more convenient to use the covariant
form for $R^{SS_z}_{\lambda_1\lambda_2}$ \cite{Jaus1,Jausa}:
 \be
        R^{SS_z}_{\lambda_1\lambda_2}(x,k_\bot)
                =\frac{1}{\sqrt{2} M_0}
        ~\bar u(p_1,\lambda_1)\Gamma v(p_2,\lambda_2), \label{covariant}
 \en
where
 \be
        &&\Gamma=\gamma_5 \qquad\qquad\qquad\qquad\qquad({\rm pseudoscalar}, S=0), \\
        &&\Gamma=-\not{\! \epsilon}(S_z)+\frac{\epsilon \cdot (p_1-p_2)}{M_0+2 m} \quad~ ({\rm vector},
        S=1).
 \en
All details are shown in Appendix A of Ref. \cite{CCH2}. The
function $H_{P,V}$ and the parameter $W_V$ are reduced to $h_{P,V}$
and $w_V$, respectively, and they are written by
 \be \label{eq:htowf}
 h_P&=&h_V=(M^2-M_0^2)\sqrt{\frac{x(1- x)}{N_c}}\frac{1}{\sqrt 2
      M_0}\phi(x,p_\bot), \non \\
 w_V&=&M_0+2m.
 \en
In principle, $\phi(x_2,p_\bot)$ is obtained by solving the
light-front QCD bound-state equation
$H_{LF}|\Psi\rangle=M|\Psi\rangle$, which is the familiar Schr{\"
o}dinger equation in ordinary quantum mechanics and $H_{LF}$ is the
light-front Hamiltonian. However, except in some simple cases, the
full solution has remained a challenge. There are several popular
phenomenological light-front momentum distribution amplitudes that
have been employed to describe various hadronic structures in the
literature. A widely used one is the Gaussian-type chosen here for
the $1S$ heavy quarkonium state \cite{hwangwei}:
 \be
 \phi^{1S}(x,p_\perp)=4 \left(\frac{\pi}{\beta^2}\right)^{3/4}
 \sqrt{\frac{dp_z}{dx}}~{\rm exp}
 \left(-\frac{p^2_z+p^2_\bot}{2
 \beta^2}\right). \label{gaussian1S}
 \en
In addition, the momentum distribution amplitude for the $2S$ heavy
quarkonium state is considered. Eqs. (\ref{eq:Norm1}) and
(\ref{gaussian1S}) can be rewritten as
 \be
 &&\int \frac{d^3 \vec{p}}{2 (2\pi)^3}
   |\phi(p)|^2 =1,\\
&& \phi^{1S}(p)=4 \left(\frac{\pi}{\beta^2}\right)^{3/4}
 {\rm exp}
 \left(-\frac{p^2}{2
 \beta^2}\right), \label{gaussian1S_p3}
 \en
where $p=|\vec{p}|$. The Fourier transition, or the conjugate
coordinate wave function of Eq. (\ref{gaussian1S_p3}) is the
ground-state solution of the harmonic-oscillator (HO) problem:
 \be
 \varphi^{1S}(r)=\left(\frac{\beta^2}{\pi}\right)^{3/4}
 {\rm exp}
 \left(-\frac{\beta^2 r^2}{2}\right). \label{coordinate1S}
 \en
Then, for consistency, the relevant excited-state solutions
 \be
 \varphi^{1P}_m(r)&=&\sqrt{\frac{8}{3}}\frac{\beta^{3/2}}{\pi^{1/4}}~\beta r~{\rm exp}
 \left(-\frac{\beta^2 r^2}{2}\right)Y_{1m}(\theta,\varphi),\label{coordinate1P} \\
 \varphi^{2S}(r)&=&\sqrt{\frac{1}{6}}\left(\frac{\beta^2}{\pi}\right)^{3/4}
 \left(\frac{3}{4}-\frac{\beta^2 r^2}{2}\right){\rm exp}
 \left(-\frac{\beta^2 r^2}{2}\right), \label{coordinate2S}
 \en
($Y_{1m}$'s are the spherical harmonics), are applied to the excited
meson states. This suggestion has been given by the authors of Ref.
\cite{ISGW}. Therefore, the Fourier transforms of Eqs.
(\ref{coordinate1P}) and (\ref{coordinate2S}) can be rewritten as
 \be
 \phi^{1P}_m(x,p_\perp)&=&4 \sqrt{2}
 \left(\frac{\pi}{\beta^2}\right)^{3/4}
 \sqrt{\frac{dp_z}{dx}}~\frac{p_m}{\beta}~{\rm exp}
 \left(-\frac{p^2_z+p^2_\bot}{2
 \beta^2}\right), \label{gaussian1P} \\
 \phi^{2S}(x,p_\perp)&=&4\sqrt{\frac{8}{3}}\left(\frac{\pi}{\beta^2}\right)^{3/4}
 \sqrt{\frac{dp_z}{dx}}~\left(\frac{p^2_z+p^2_\perp}{2\beta^2}-\frac{3}{4}\right){\rm exp}
 \left(-\frac{p^2_z+p^2_\bot}{2
 \beta^2}\right), \label{gaussian2S}
 \en
($p_{m=\pm1}=\mp (p_{\perp 1}\pm i p_{\perp 2})/\sqrt{2}$, and
$p_{m=0}=p_z$) and can be treated as the momentum distribution
amplitudes of the $1P$ and $2S$ meson states, respectively. In fact,
Eq. (\ref{gaussian1P}) has been used for the $p$-wave mesons in Ref.
\cite{CCH2}. This paper further analyzes the momentum distribution
amplitudes as shown in Eqs. (\ref{gaussian1S}) and
(\ref{gaussian2S}).

The formulae of the decay constants $f_{P,V}$ and M1 transition form
factor $V(0)$ are needed in the latter analysis. However, their
derivations have been done in Refs. \cite{CCH2,hwangwei}. The
definitions and formulae for them are provided here.

\subsection{Decay constant $f_{P,V}$}
The decay constants of mesons $f_{P,V}$ are defined by the
matrix elements for $P$ and $V$ mesons
 \be
 \la 0|A_\mu |P(P) \ra &=& if_P P_\mu, \\
 \la 0|V_\mu |V(P,\epsilon) \ra &=& M_V f_V \epsilon_\mu,
 \en
where $P_\mu$ is the momentum of meson and $\epsilon_\mu$ is the
polarization vector of $V$ meson. The formula for $f_P$ is
 \be
 f_P &=& \frac{N_c}{4\pi^3}\int dx d^2p_\bot \frac{h_P}{x(1-x)
          (M^{2}-M^{2}_0)}~m \non \\
     &=& \frac{\sqrt{2N_c}}{8\pi^3}\int dx d^2 p_\t
 \frac{m}{\sqrt{m^2+p_\t^2}}\phi_P(x,p_{\bot}), \label{eq:dcP}
 \en
where $N_c=3$ is the color number and $m$ denotes the quark mass. As
to the formula for $f_V$, we considered the case with the transverse
polarization
 \be
 \epsilon(\pm)=\left(\frac{2}{P^+}\epsilon_\perp \cdot
 P_\perp,0,\epsilon_\perp\right),~~~\epsilon_\perp =
 \mp\frac{1}{\sqrt{2}}(1,\pm i).
 \en
and obtain
 \be
 f_V &=&\frac{N_c}{4\pi^3}\int dx d^2p_\bot \frac{h_V}{x_1
         x_2 (M^{2}-M^{2}_0)}\left[x M^{2}_0-p^{2}_\bot
         +\frac{2 m}{W_V}\,p^{2}_\bot
         \right]\non \\
            &=& \frac{\sqrt{2N_c}}{8\pi^3 M}\int dx d^2 p_\t
     \frac{1}{\sqrt{m^2+p_\t^2}}\left [\frac{M_0^2}{2}-
     p_\t^2+\frac{2m}{w_V}p_\t^2 \right ]\phi_V(x,p_{\bot}). \label{eq:dcV}
 \en
This expression can be shown to be in agreement with Eq. (2.22) of
Ref. \cite{CCH2}. Since the momentum distribution function is even
in $p_z$, a quality defined in Eq. (\ref{eq:Mpz}), it follows that
 \be
 \int dx d^2p_\perp
 \frac{\phi_V(x,p_\perp)}{\sqrt{m^2+p^2_\perp}}\left(x-\frac{1}{2}\right)M_0=0.
 \en
Therefore,
 \be
 \int dx d^2p_\perp
 \frac{\phi_V(x,p_\perp)}{\sqrt{m^2+p^2_\perp}}~xM_0^2=\int dx d^2p_\perp
 \frac{\phi_V(x,p_\perp)}{\sqrt{m^2+p^2_\perp}}\frac{M_0^2}{2}.
 \en

\subsection{Vector current form factor $V(q^2)$}

For the transition $V\to P\gamma$, a more general process $V\to
P\gamma^*$ where the final photon is off-shell is considered. The
$V\to P\gamma^*$ transition is parameterized in term of a vector
current form factor $V(q^2)$ by
 \be
 \Gamma_{\mu}=ie\varepsilon_{\mu\nu\alpha\beta}\epsilon^{\nu}q^{\alpha}P^{\beta}V(q^2),
 \en
where $\Gamma_{\mu}$ is the amplitude of the $V\to P\gamma^*$
process. $P$ ($\epsilon$) is the momentum (polarization vector) of
the initial vector meson, $P'$ denotes the momentum of the final
pseudoscalar meson, and the momentum transfer $q=P-P'$. The formula
for the form factor $V(q^2)$ is
 \be \label{eq:Vq2}
 V(q^2)&=&\frac{e_q}{8\pi^3}\int dx d^2p_{\bot}
   \frac{\phi_V(x,p_\bot)}{M_0}\left[\frac
   {\phi_P(x,p'_\bot)}{(1-x)M'_0}+\frac{\phi_P(x,p''_\bot)}{x
   M''_0}\right]\non \\
   &&\qquad\qquad\qquad\times\left[ m+\frac{2}{w_V}\left( p_{\bot}^2+
    \frac{(p_{\bot}\cdot q_{\bot})^2}{q^2} \right ) \right ],
 \en
where $p'_\perp=p_\perp -(1-x) q_\perp$, $p''_\perp=p_\perp +x
q_\perp$, $M'^2_0=(m^2+p'^2_\perp)/x (1-x)$, and
$M''^2_0=(m^2+p''^2_\perp)/x (1-x)$. The rate for $V\to P\gamma$ is
 \be \label{eq:VPr2}
 \Gamma(V\to P\gamma)=\frac{\alpha}{3}\frac{(M_V^2-M_P^2)^3}{8 M_V^3}|V(0)|^2.
 \en

\section{Analysis of momentum distribution amplitude}
In this section, the momentum distribution amplitudes of
pseudoscalar and vector heavy quarkonium states are analyzed. The
forms of quark distribution amplitude and $\xi$-moments can be
derived from them.
\subsection{Quark distribution amplitude $\Phi_P(\xi)$}
The quark distribution amplitude of the pseudoscalar heavy
quarkonium state can be defined as follows \cite{CZ}:
 \be
 \langle 0|\bar c(z)\gamma^\alpha \gamma_5[z,-z]c(-z)|P\rangle=if_P
 P^\alpha\int^1_{-1}d\xi e^{i(Pz)\xi}\Phi(\xi,\mu),\label{LCWF}
 \en
where $\xi=2x-1$ and $\mu$ is an energy scale which separates the
perturbative and non-perturbative regimes. The factor $[z,-z]$ is
defined as
 \be
 [z,-z]=P{\rm exp}[iq\int^z_{-z}dx^\mu A_\mu(x)],
 \en
which makes the matrix element Eq. (\ref{LCWF}) gauge invariant. The
quark distribution amplitude $\Phi(\xi,\mu)$ is normalized as
 \be
 \int^1_{-1}d\xi \Phi(\xi,\mu)=1,
 \en
and it can be expanded \cite{CZ} in Gegenbauer polynomials
$C^{3/2}_n(\xi)$ as
 \be
 \Phi(\xi,\mu)= \Phi_{as}(\xi)\left[1+\sum_{n=1}^\infty
 a_n(\mu)C_n^{3/2}(\xi)\right],
 \en
where $\Phi_{as}(\xi)=3(1-\xi^2)/4$ is the asymptotic quark
distribution amplitude and $a_n(\mu)$ the Gegenbauer moments which
describe to what degree the quark distribution amplitude deviates
from the asymptotic one. $C^{3/2}_n(\xi)$s have the orthogonality
integrals
 \be
 \int^1_{-1} (1-\xi^2)C^{3/2}_m(\xi) C^{3/2}_n (\xi) d\xi =
 \frac{2(n+1)(n+2)}{2n+3}~\delta_{mn}. \label{orthogonal}
 \en
Then $a_n$ can be obtained by using the above orthogonality
integrals as
 \be
 a_n(\mu) &=&\frac{2(2
 n+3)}{3(n+1)(n+2)}\int^1_{-1}C^{3/2}_n(\xi)\Phi(\xi,\mu)d\xi.
 \en
An alternative approach to parameterize quark distribution amplitude
is to calculate the so-called $\xi$-moments,
 \be
 \langle \xi^n\rangle_\mu=\int^1_{-1} d\xi~\xi^n \Phi(\xi,\mu).
 \en
It is easy to find the relations between the Gegenbauer moments and
the $\xi$-moments. Here we list them up to $n=6$
 \be
 a_2 &=& -\frac{7}{12}[1-5\langle
 \xi^2\rangle],\non \\
 a_4 &=& \frac{11}{24}[1-14\langle
 \xi^2\rangle + 21\langle \xi^4\rangle],\non \\
 a_6 &=& -\frac{35}{448}[5-135\langle
 \xi^2\rangle + 495\langle \xi^4\rangle-429 \langle
 \xi^6\rangle].\label{anmoments}
 \en
All the $n$-odd moments are vanishing because this paper only
focuses on the $s$-wave momentum distribution amplitudes, which are
$\xi$-even functions.

There are some similar procedures \cite{BHL,DM} by which the quark
distribution amplitude can be obtained from the equal time wave
function. Within the framework of this study, the decay constant Eq.
(\ref{eq:dcP}) can be rewritten as
 \be
 1=\int^1_0 dx \left[\frac{\sqrt{2N_c}}{8\pi^3~f_P}\int d^2p_\perp
 \frac{m}{\sqrt{m^2+p_\t^2}}\phi_P(x,p_{\bot})\right]
 \en
and we defined LCWF $\hat{\Phi}_P(x,\mu)$ as
 \be
 \hat{\Phi}_P(x,\mu)=\frac{\sqrt{2N_c}}{8\pi^3~f_P}\int^{p_\perp^2 < \mu^2} d^2p_\perp
 \frac{m}{\sqrt{m^2+p_\t^2}}\phi_P(x,p_{\bot})
 \en
where $\hat{\Phi}_P(x,\mu)=2 \Phi_P(\xi,\mu)$. If the momentum
distribution amplitudes Eqs. (\ref{gaussian1S}) and
(\ref{gaussian2S}) are applied as the non-perturbative inputs, the
suppression of the Gaussian function allows us to do the $p_\perp$
integrals up to infinity with no loss of accuracy. Thus, the results
can be obtained as follows:
 \be
 \Phi^{1S}_P(\xi)&=&\sqrt{\frac{3}{8}} \left(\frac{2}{\pi}\right)^{5/4}
 \frac{m}{f^{1S}_P}~e^d~\Gamma\left[\frac{3}{4}, \frac{d}{1-\xi^2}\right],\label{LC1S}\\
 \Phi^{2S}_P(\xi)&=&\left(\frac{2}{\pi}\right)^{5/4}
 \frac{m}{f^{2S}_P}~e^d~\left\{\Gamma\left[\frac{7}{4},\frac{d}{1-\xi^2}\right]
 -\left(\frac{3}{4}+d\right)~
 \Gamma\left[\frac{3}{4},\frac{d}{1-\xi^2}\right]\right\},\label{LC2S}
 \en
where $\Gamma[r,y]$ is the incomplete Gamma function
 \be
 \Gamma[r,y]=\int^\infty_y t^{r-1}e^{-t} dt, \label{gamma}
 \en
and $d=m^2/2 \beta^2$. The incomplete gamma function may be
expressed quite elegantly in terms of the confluent hypergeometric
function
 \be
 \Gamma[r,y]=\Gamma[r]-r^{-1}y^r\times~_1F_1(r;r+1;-y),\label{hpgeo}
 \en
where
 \be
 _iF_j(r_1,r_2,...,r_i;r'_1,r'_2,...,r'_j;y)=\sum^\infty_{n=0}\frac{(r_1)_n(r_2)_n
 ...(r_i)_n}{(r'_1)_n(r'_2)_n...(r'_j)_n}\frac{y^n}{n!},\label{1F1}
 \en
and $(r)_n=(r+n-1)!/(r-1)!$ is the Pochhammer symbol. 
After fixing the parameters $m$ and $\beta$, the behaviors of the
quark distribution amplitudes, Eqs. (\ref{LC1S}) and (\ref{LC2S}),
are determined and compared with those of other theoretical groups
in Section IV. It is worth to mention that Eqs. (\ref{LC1S}) and
(\ref{LC2S}) have incorrect asymptotic behavior because the
Gaussian-type wave functions are primarily only suitable for the low
energy region. At energy scales accessible at current experiments, a
quark distribution amplitude can be far from the asymptotic form.
The worth of the asymptotic form is that it provides a criterion. In
other words, one can use the Gegenbauer moments (or $\xi$-moments)
to describe to what degree his quark distribution amplitude deviates
from the asymptotic one.

In fact, the $\xi$ integrals of $\Phi^{1(2)S}_P(\xi)$ can also be
analytically performed, and the decay constants $f_P^{1(2)S}$ can be
expressed as
 \be
 f_P^{1S} &=& \sqrt{\frac{3}{2}} \left(\frac{2}{\pi}\right)^{5/4}m~
 e^d \left\{\Gamma\left[\frac{3}{4}\right]~_1F_1\left(-\frac{1}{2}; \frac{1}{4};
 -d\right)
 - \frac{3d^{3/4} \Gamma[-\frac{3}{4}]^2}{8\sqrt{2\pi}}~_1F_1\left(\frac{1}{4};
\frac{7}{4}; -d\right)\right\}, \label{1Sanalytic}\\
 f_P^{2S} &=& 3\left(\frac{2}{\pi}\right)^{5/4}
 m~e^d \left\{-2 d~ \Gamma\Bigg[\frac{3}{4}\right]~_1F_1\left(-\frac{1}{2}; \frac{5}{4}; -d\right)
 +\frac{3 d^{3/4} \Gamma[-\frac{3}{4}]^2}{16\sqrt{2\pi}}
  ~_1F_1\left(-\frac{3}{4}; \frac{3}{4}; -d\right)\Bigg\}.\label{2Sanalytic}
 \en
It is evident that $f_P^{1(2)S}$ is only dependent on the values of
$m$ and $m/\beta$. Eqs. (\ref{1Sanalytic}) and (\ref{2Sanalytic})
can be expanded in terms of $d$ as
 \be
 f_P^{1S} &\sim&
 c_1\left(1+3d+\frac{21}{10}d^2+\frac{77}{90}d^3+...\right)+c'_1 d^{3/4}
 \left(1+\frac{6}{7}d+\frac{30}{77}d^2+\frac{4}{33}d^3+...\right),\label{1Sseries}\\
 f_P^{2S} &\sim&-c_2d\left(1+\frac{7}{5}d+\frac{77}{90}d^2+\frac{77}{234}d^3...\right)+
 c'_2 d^{3/4}\left(1+2 d+\frac{10}{7}d^2+\frac{20}{33}d^3+...\right),\label{2Sseries}
 \en
where all $c^{(_{'})}_{1,2}$ are positive constants. However, for a
typical value of $d$, all the series in Eqs. (\ref{1Sseries}) and
(\ref{2Sseries}) converge very slowly. In addition, the $\xi$-
moments can be expressed analytically as
 \be
 \langle \xi^2\rangle^{1S}_P &=& A^{1S}_P\Bigg\{\frac{2}{3} \Gamma\left[\frac{3}{4}\right]
 ~_1F_1\left(-\frac{3}{2};\frac{1}{4}, -d\right)-\frac{d^{3/4}  \Gamma\left[-\frac{3}{4}
 \right]^2}{2 \sqrt{2\pi}}~_1F_1\left(-\frac{3}{4}, \frac{7}{4},
 -d\right)\bigg\},\non \\
 \langle \xi^4\rangle^{1S}_P &=& A^{1S}_P\Bigg\{\frac{2}{5} \Gamma\left[\frac{3}{4}\right]
 ~_1F_1\left(-\frac{5}{2};\frac{1}{4}, -d\right)-\frac{3d^{3/4}  \Gamma\left[-\frac{3}{4}
 \right]^2}{7 \sqrt{2\pi}}~_1F_1\left(-\frac{7}{4}, \frac{7}{4},
 -d\right)\bigg\},\non \\
 \langle \xi^6\rangle^{1S}_P &=& A^{1S}_P\Bigg\{\frac{2}{7}
 \Gamma\left[\frac{3}{4}\right]
 ~_1F_1\left(-\frac{7}{2};\frac{1}{4}, -d\right)-\frac{30d^{3/4}  \Gamma\left[-\frac{3}{4}
 \right]^2}{77 \sqrt{2\pi}}~_1F_1\left[-\frac{11}{4}, \frac{7}{4},
 -d\right]\bigg\},\label{ximoment1s}
 \en
and
 \be
  \langle \xi^2\rangle^{2S}_P &=&
  A^{2S}_P\Bigg\{\frac{7d}{6}\Gamma\left[-\frac{1}{4}\right]~_1F_1\left(-\frac{3}{2},\frac{5}{4};-d\right)
  +\frac{3d^{3/4}\Gamma\left[-\frac{3}{4}\right]^2}{8\sqrt{2\pi}}
  ~_1F_1\left(-\frac{7}{4},\frac{3}{4};-d\right)
  \Bigg\},\non \\
 \langle \xi^4\rangle^{2S}_P &=&
  A^{2S}_P\Bigg\{\frac{11d}{10}\Gamma\left[-\frac{1}{4}\right]~_1F_1\left(-\frac{5}{2},\frac{5}{4};-d\right)
  +\frac{9d^{3/4}\Gamma\left[-\frac{3}{4}\right]^2}{28\sqrt{2\pi}}
  ~_1F_1\left(-\frac{11}{4},\frac{3}{4};-d\right)
  \Bigg\},\non \\
  \langle \xi^6\rangle^{2S}_P &=&
  A^{2S}_P\Bigg\{\frac{15d}{14}\Gamma\left[-\frac{1}{4}\right]~_1F_1\left(-\frac{7}{2},\frac{5}{4};-d\right)
 +\frac{90d^{3/4}\Gamma\left[-\frac{3}{4}\right]^2}{308\sqrt{2\pi}}
  ~_1F_1\left(-\frac{15}{4},\frac{3}{4};-d\right)
  \Bigg\},
  \label{ximoment2s}
 \en
where
 \be
 A^{1S}_P = \sqrt{\frac{3}{8}} \left(\frac{2}{\pi}\right)^{5/4}
 \frac{m}{f^{1S}_P}~e^d,\quad\quad
 A^{2S}_P = \left(\frac{2}{\pi}\right)^{5/4}
 \frac{m}{f^{2S}_P}~e^d.
 \en
The derivations of Eqs. (\ref{ximoment1s}) and (\ref{ximoment2s})
have used the formula
 \be
 _1
 F_1(a;b;c)=\frac{b-1}{c}\left[~_1F_1(a;b-1;c)-~_1F_1(a-1;b-1;c)\right],
 \en
which is easily checked from the definition of the confluent
hypergeometric function Eq. (\ref{1F1}).

\subsection{Quark distribution amplitude $\Phi_V(\xi)$}
Similar to the case of $\Phi_P(\xi)$, Eq. (\ref{eq:dcV}) can be
rewritten and $\hat{\Phi}_V(x)$ defined as
 \be
 \hat{\Phi}_V(x) = \frac{\sqrt{2N_c}}{8\pi^3 M~f_V}\int d^2 p_\t
     \frac{\phi_V(x,p_{\bot})}{\sqrt{m^2+p_\t^2}}\left [\frac{M_0^2}{2}-
     p_\t^2+\frac{2m}{w_V}p_\t^2 \right ]. \label{LCWFvector}
 \en
However, the $p_\perp$ integrals in Eq. (\ref{LCWFvector}) cannot be
analytically performed . The crux is the third term in the square
bracket, that is, the term proportional to $1/w_V$. This term may be
rewritten and expanded as
 \be \label{vwapp}
 \frac{1}{M_0+2m}=\frac{1}{4m\left(1+\frac{M_0-2m}{4m}\right)}
 \simeq
 \frac{1}{4m}\left[1-\left(\frac{M_0-2m}{4m}\right)+\left(\frac{M_0-2m}{4m}\right)^2-...\right],
 \en
because, in the non-relativistic limit, $M_0 \to 2m$. Then, we
defined the approximate quark distribution amplitudes as
 \be
 &&\hat{\Phi}'_V(x) = \frac{\sqrt{2N_c}}{8\pi^3 M~f_V}\int d^2 p_\t
     \frac{\phi_V(x,p_{\bot})}{\sqrt{m^2+p_\t^2}}\left [\frac{M_0^2}{2}-
     p_\t^2+\frac{p_\t^2}{2}\right], \label{PhiV0}\\
 &&{\hat\Phi}''_V(x) = \frac{\sqrt{2N_c}}{8\pi^3 M~f_V}\int d^2 p_\t
     \frac{\phi_V(x,p_{\bot})}{\sqrt{m^2+p_\t^2}}\left [\frac{M_0^2}{2}-
     p_\t^2+\frac{p_\t^2}{2}\left(1-\frac{M_0-2m}{4m}\right) \right
     ].\label{PhiV1}
 \en
After applying the momentum distribution function in Eq.
(\ref{gaussian1S}) and fitting the parameters, we found that not
only the center values of the parameters $\beta$ of Eqs.
(\ref{PhiV0}) and (\ref{PhiV1}) were both in the error bar of
$\beta_{J/\psi}$ (see Table I), but also that their curves were
almost the same. In addition, a similar situation also existed in
the case of the $2S$ states. Therefore, we have only shown the
results from the form $\hat{\Phi}'_V(x)$. By performing the
$p_\perp$ integrals, it can be obtained that
 \be
 \Phi'^{1S}_V(\xi) &=& \sqrt{\frac{3}{8}}\left(\frac{2}{\pi}\right)^{5/4}\frac{w^2e^d}{Mf_V^{1S}}
 \Bigg\{d\Gamma\left[\frac{3}{4},\frac{d}{1-\xi^2}\right]+(3+\xi^2)
 \Gamma\left[\frac{7}{4},\frac{d}{1-\xi^2}\right]\bigg\}, \label{analyPhiV1s}  \\
 {\hat\Phi}'^{2S}_V(\xi) &=& \left(\frac{2}{\pi}\right)^{5/4}\frac{w^2e^d}{Mf_V^{2S}}
 \Bigg\{d\Gamma\left[\frac{7}{4},\frac{d}{1-\xi^2}\right]+(3+\xi^2)
 \Gamma\left[\frac{11}{4},\frac{d}{1-\xi^2}\right]\non \\
 &&\qquad\qquad\qquad\quad -\left(\frac{3}{4}+d\right)\left(d\Gamma\left[\frac{3}{4},
 \frac{d}{1-\xi^2}\right]+(3+\xi^2)\Gamma\left[\frac{7}{4},\frac{d}{1-\xi^2}\right]\right)\bigg\}.
 \label{analyPhiV2s}
 \en
The above derivations are based on the formula:
 \be
 \Gamma[r,y]-(r-1)\Gamma[r-1,y]=y(\Gamma[r-1,y]-(r-2)\Gamma[r-2,y]),
 \en
which is easily checked from the definition of incomplete Gamma
function, Eq. (\ref{gamma}). The $\xi$-moments of $\Phi'_V(\xi)$ can
also be obtained analytically
 \be
 \langle \xi^2\rangle^{1S}_V &=& A^{1S}_V\Bigg\{\frac{1}{5} \Gamma\left[\frac{3}{4}\right]
 \left[10~_1F_1\left(-\frac{3}{2};-\frac{3}{4};
 -d\right)-~_1F_1\left(-\frac{5}{2};-\frac{3}{4};
 -d\right)\right] \non \\
 &&\qquad+\frac{2d^{7/4}  \Gamma\left[-\frac{3}{4}
 \right]^2}{7 \sqrt{2\pi}}\bigg[3~_1F_1\left(\frac{1}{4};
 \frac{11}{4};
 -d\right)-~_1F_1\left(-\frac{3}{4};
 \frac{11}{4};
 -d\right)\bigg]\bigg\},\non \\
 \langle \xi^4\rangle^{1S}_V &=& A^{1S}_V\Bigg\{\frac{3}{35} \Gamma\left[\frac{3}{4}\right]
 \left[14~_1F_1\left(-\frac{5}{2};-\frac{3}{4};
 -d\right)-~_1F_1\left(-\frac{7}{2};-\frac{3}{4};
 -d\right)\right] \non \\
 &&\qquad+\frac{3 d^{7/4}  \Gamma\left[-\frac{7}{4}
 \right]^2}{4 \sqrt{2\pi}}\bigg[7~_1F_1\left(-\frac{3}{4};
 \frac{11}{4};
 -d\right)-~_1F_1\left(-\frac{7}{4};
 \frac{11}{4};
 -d\right)\bigg]\bigg\},\non
 \en
 \be
 \langle \xi^6\rangle^{1S}_V &=& A^{1S}_V\Bigg\{\frac{1}{21} \Gamma\left[\frac{3}{4}\right]
 \left[18~_1F_1\left(-\frac{7}{2};-\frac{3}{4};
 -d\right)-~_1F_1\left(-\frac{9}{2};-\frac{3}{4};
 -d\right)\right] \non \\
 &&\qquad+\frac{165 d^{7/4}  \Gamma\left[-\frac{11}{4}
 \right]^2}{32 \sqrt{2\pi}}\bigg[11~_1F_1\left(-\frac{7}{4};
 \frac{11}{4};
 -d\right)-~_1F_1\left(-\frac{11}{4};
 \frac{11}{4};
 -d\right) \bigg]\bigg\},\label{ximomentV1s}
 \en
and
 \be
 \langle \xi^2\rangle^{2S}_V &=& A^{2S}_V\Bigg\{\frac{1}{5} \Gamma\left[\frac{3}{4}\right]
 \left[15~_1F_1\left(-\frac{3}{2};-\frac{3}{4};
 -d\right)+(d-6)~_1F_1\left(-\frac{5}{2};-\frac{3}{4};
 -d\right)\right] \non \\
 &&\qquad+\frac{d^{7/4}  \Gamma\left[-\frac{3}{4}
 \right]^2}{7 \sqrt{2\pi}}\bigg[9~_1F_1\left(\frac{1}{4};
 \frac{11}{4};
 -d\right)+2(d-6)~_1F_1\left(-\frac{3}{4};
 \frac{11}{4};
 -d\right)\bigg]\bigg\},\non \\
 \langle \xi^4\rangle^{2S}_V &=& A^{2S}_V\Bigg\{\frac{3}{35} \Gamma\left[\frac{3}{4}\right]
 \left[35~_1F_1\left(-\frac{5}{2};-\frac{3}{4};
 -d\right)+(d-22)~_1F_1\left(-\frac{7}{2};-\frac{3}{4};
 -d\right)\right] \non \\
 &&\quad+\frac{3 d^{7/4}  \Gamma\left[-\frac{7}{4}
 \right]^2}{8 \sqrt{2\pi}}\bigg[35~_1F_1\left(-\frac{3}{4};
 \frac{11}{4};
 -d\right)+2(d-22)~_1F_1\left(-\frac{7}{4};
 \frac{11}{4};
 -d\right)\bigg]\bigg\},\non \\
 \langle \xi^6\rangle^{2S}_V &=& A^{2S}_V\Bigg\{\frac{1}{21} \Gamma\left[\frac{3}{4}\right]
 \left[63~_1F_1\left(-\frac{7}{2};-\frac{3}{4};
 -d\right)+(d-46)~_1F_1\left(-\frac{9}{2};-\frac{3}{4};
 -d\right)\right] \non \\
 &&\quad+\frac{165 d^{7/4}  \Gamma\left[-\frac{11}{4}
 \right]^2}{64 \sqrt{2\pi}}\bigg[77~_1F_1\left(-\frac{7}{4};
 \frac{11}{4};
 -d\right)+2 (d-46)~_1F_1\left(-\frac{11}{4};
 \frac{11}{4};
 -d\right) \bigg]\bigg\},\label{ximomentV2s}
 \en
where
 \be
 A^{1S}_V=\sqrt{\frac{3}{8}}\left(\frac{2}{\pi}\right)^{5/4}\frac{w^2e^d}{Mf_V^{1S}},\quad
 A^{2S}_V=\left(\frac{2}{\pi}\right)^{5/4}\frac{w^2e^d}{Mf_V^{2S}}.
 \en

\section{Numerical results and discussions}
In this work, the numerical results were calculated for the quark
distribution amplitudes $\Phi(\xi)$ and $\xi$-moments for the
pseudoscalar and vector heavy quarkonium. First, it was necessary to
determine the parameters appearing in the momentum distribution
functions. In total, there are five for charmonium, $m_c$,
$\beta_{\eta_c}$, $\beta_{\eta'_c}$, $\beta_{J/\psi}$,
$\beta_{\psi'}$, and five ones for bottomonium: $m_b$,
$\beta_{\eta_b}$, $\beta_{\eta'_b}$, $\beta_{\Upsilon}$,
$\beta_{\Upsilon'}$. The decay constants of vector heavy quarkonium
states were determined first. The decay constant $f_V$ is related to
the electromagnetic decay of vector meson $V\to e^+e^-$ by \cite{NS}
 \be \label{eq:dcVexp}
 \Gamma(V\to e^+e^-)=\frac{4\pi}{3}\frac{\alpha^2}{M_V}c_V f_V^2,
 \en
where $c_V$ is the square of the electric charge of heavy quark.
From the data \cite{PDG08}, we obtained the values
 \be
 f_{J/\psi}&=&416 \pm 7~{\rm MeV},\qquad\qquad f_{\psi'}=298 \pm 8~{\rm
 MeV}, \label{fvectorc}\\
 f_{\Upsilon}&=&715 \pm 5~{\rm MeV},\qquad\qquad f_{\Upsilon'}=497 \pm 4~{\rm
 MeV}.\label{fvectorb}
 \en
For the decay constants of $\eta_c$, the two decay modes $B\to
K\eta_c$ and $B\to K J/\psi$ are considered using the following
relation
 \be
 \frac{\Gamma(B\to K \eta_c)}{\Gamma(B\to K J/\psi)}=
 \left(\frac{f_{\eta_c}}{f_{J/\psi}}\right)^2
 \Bigg|\frac{C_{\eta_c}}{C_{J/\psi}}\Bigg|^2\frac{\lambda_{BK\eta_c}}{\lambda_{BK J/\psi}^3}
 \Bigg|\left(1-\frac{M_K^2}{M_B^2}\right)\frac{f_+(M_{\eta_c}^2)}{f_+(M_{J/\psi}^2)}+
 \frac{M_{\eta_c}^2}{M_B^2}\frac{f_-(M_{\eta_c}^2)}{f_+(M_{J/\psi}^2)}\Bigg|^2,\label{Btoccratio}
 \en
where $C_{\eta_c}$ and $C_{J/\psi}$ are related to the Wilson
coefficients and can be determined to have the value
$|C_{\eta_c}/C_{J/\psi}|=0.89\pm0.02$ \cite{DT},
 \be
 \lambda_{abc}=\left[\left(1-\frac{M_b^2}{M_a^2}-\frac{M_c^2}{M_a^2}\right)^2-
 4\frac{M_b^2}{M_a^2}\frac{M_c^2}{M_a^2}\right]^{1/2},
 \en
and the form factors $f_\pm(q^2)$ are defined by the Lorentz
decomposition of the matrix element
 \be
 \langle K(p)|\bar s \gamma_\mu b | B(p+q)\rangle=f_+(q^2)
 (2p+q)_\mu + f_-(q^2)q_\mu.
 \en
By the results of Ref. \cite{CCH2} one calculated the form factors
$f_\pm(q^2)$ within the covariant light-front quark model. Combining
the above with the experimental values, we can obtain
 \be
 f_{\eta_c}=421 \pm 38~{\rm MeV}. \label{fetac}
 \en
This result leads to the ratio
 \be
 \left(\frac{f_{\eta_c}}{f_{J/\psi}}\right)^2=1.02\pm 0.22,\label{thisworkratio}
 \en
which is consistent with the Van Royen-Weisskopf formula \cite{VRW},
 \be
 \left(\frac{f_{\eta_c}}{f_{J/\psi}}\right)^2=\frac{M_{J/\psi}}
 {M_{\eta_c}}\frac{|\Psi_{\eta_c}(0)|^2}{|\Psi_{J/\psi}(0)|^2}\simeq
 1.04,\label{VRWratio}
 \en
when the approximation $\Psi_{\eta_c}(0)\simeq \Psi_{J/\psi}(0)$ is
used. For the decay constant $f_{\eta'_c}$, however, a similar
estimation of Eq. (\ref{Btoccratio}) has a large uncertainty because
the error of $Br(B\to K \eta'_c)$ was greater than $50\%$
\cite{PDG08}. Given the consistency between Eq.
(\ref{thisworkratio}) and (\ref{VRWratio}), the approximation
$\Psi_{\eta'_c}(0)\simeq \Psi_{\psi'}(0)$ was used, and the
corresponding Van Royen-Weisskopf formula,
 \be
 \left(\frac{f_{\eta'_c}}{f_{\psi'}}\right)^2\simeq\frac{M_{\psi'}}
 {M_{\eta'_c}}= 1.01, \label{appratio}
 \en
is applied and the decay constant $f_{\eta'_c}=300\pm 8$MeV is
obtained. Furthermore, an additional constraint is needed. The new
data $Br(J/\psi \to \eta_c \gamma)=(1.98\pm0.31)\%$ \cite{newdata}
are chosen to calculate the form factor $V(0)$ in Eq.
(\ref{eq:VPr2}). For the charm sector, there were one assumption
(Eq. (\ref{appratio})) and four constraints (the decay constants
(\ref{fvectorc}), (\ref{fetac}), and the new data $Br(J/\psi \to
\eta_c \gamma)$) were used to fix the five parameters appearing in
the momentum distribution amplitudes. The results are listed in
Table I.
\begin{table}[h!]
\caption{\label{tab:para} Fixed parameters $m_c$ and $\beta$'s for
charmonium states (in units of GeV). $\beta'$ and $\beta''$ are the
parameters appearing in equations (\ref{PhiV0}) and (\ref{PhiV1}).}
\begin{ruledtabular}
\begin{tabular}{c|ccccc}
 parameter & $m_c$    & $\beta_{\eta_c}$ & $\beta_{\eta'_c}$ & $\beta_{J/\psi}$ &  $\beta_{\psi'}$
          \\ \hline
 value & $1.56$  & $0.820^{+0.078}_{-0.074}$     &
$0.665^{+0.028}_{-0.026}$    & $0.613 \pm 0.006$     & $0.477 \pm
0.008$ \\ \hline\hline
 parameter & & $\beta'_{J/\psi}$ & $\beta''_{J/\psi}$ & $\beta'_{\psi'}$
 & $\beta''_{\psi'}$ \\ \hline
 value & & $0.611 \pm 0.06 $ & $0.613 \pm 0.06 $ & $0.474\pm 0.08 $ & $0.477\pm 0.08 $
\end{tabular}
\end{ruledtabular}
\end{table}
As to the decay constants of $\eta_b$ and $\eta_b'$, a similar
approximation was assumed
 \be
 \left(\frac{f_{\eta'_b}}{f_{\Upsilon'}}\right)^2\simeq\left(\frac{f_{\eta_b}}
 {f_{\Upsilon}}\right)^2\simeq\frac{M_{\Upsilon}}
 {M_{\eta_b}}= 1.01, \label{appratiob}
 \en
and the decay constants $f_{\eta_b}=718\pm 5$ MeV,
$f_{\eta'_b}=499\pm 5$ MeV obtained. Finally we used \cite{CCH2}
$m_b=4.64$ GeV and fixed the $\beta$ parameters for the bottom
sector in Table II.
\begin{table}[h!]
\caption{\label{tab:parab} Parameters $m_b$ and $\beta$'s for
bottomonium states (in units of GeV). $\beta'$ and $\beta''$ are the
parameters appearing in equations (\ref{PhiV0}) and (\ref{PhiV1}).}
\begin{ruledtabular}
\begin{tabular}{c|ccccc}
 parameter & $m_b$    & $\beta_{\eta_b}$ & $\beta_{\eta'_b}$ & $\beta_{\Upsilon}$ &  $\beta_{\Upsilon'}$
          \\ \hline
 value & $4.64$  & $1.47\pm 0.01$     &
$1.04\pm 0.01$    & $1.30 \pm 0.01$     & $0.926 \pm 0.005$
\\ \hline\hline
 parameter & & $\beta'_{\Upsilon}$ & $\beta''_{\Upsilon}$ & $\beta'_{\Upsilon'}$
 & $\beta''_{\Upsilon'}$ \\ \hline
 value & & $1.30 \pm 0.01 $ & $1.30 \pm 0.01 $ & $0.924\pm 0.05 $ & $0.926\pm 0.005 $
\end{tabular}
\end{ruledtabular}
\end{table}

After fixing all parameters, Eqs. (\ref{LC1S}), (\ref{LC2S}),
(\ref{analyPhiV1s}), and (\ref{analyPhiV2s}) are used to plot the
quark distribution amplitudes $\Phi^{(_{'})1S}_{P,V}(\xi)$ and
$\Phi^{(_{'})2S}_{P,V}(\xi)$ for the charm sector, as shown in Fig.
1 and Fig. 2, respectively.
\begin{figure}
\includegraphics*[width=4in]{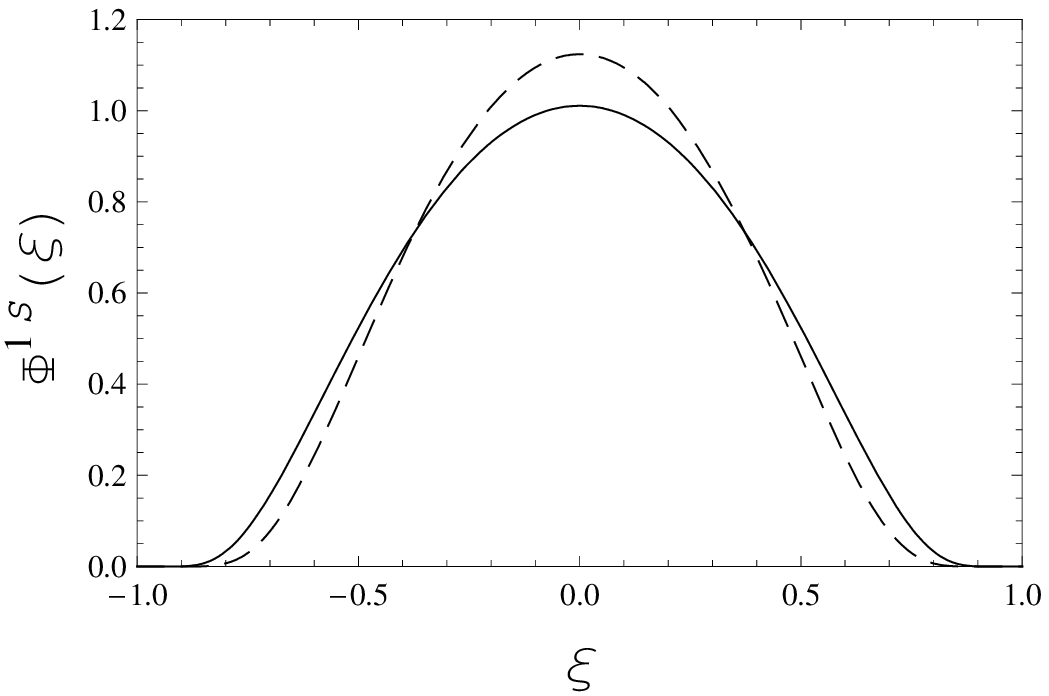}
\caption{ Quark distribution amplitudes $\Phi_{\eta_c}(\xi)$ (solid
lines) and $\Phi'_{J/\psi}(\xi)$ (dashed lines) of this work.}
 \label{fig:xi1S}
\end{figure}
\begin{figure}
\includegraphics*[width=4in]{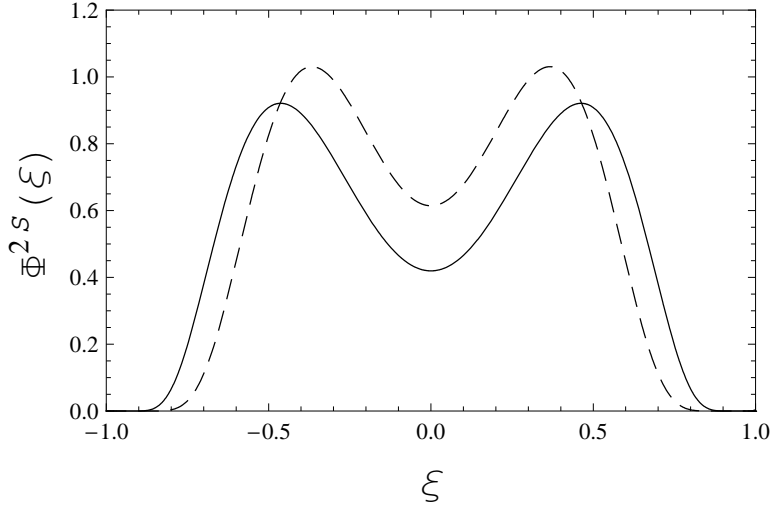}
\caption{ Quark distribution amplitudes $\Phi_{\eta'_c}(\xi)$ (solid
lines) and $\Phi'_{\psi'}(\xi)$ (dashed lines) of this work.}
 \label{fig:xi2S}
\end{figure}
As is seen, the momentum fraction $x$ of the pseudoscalar charmonium
had a slightly wider distribution than that of the vector
charmonium. Next, the quark distribution amplitudes
$\Phi_{\eta_c}(\xi)$ and $\Phi_{\eta'_c}(\xi)$ of this work were
compared with those of Ref. \cite{Braguta1,Braguta2} in Fig. 3.
\begin{figure}
\includegraphics*[width=4in]{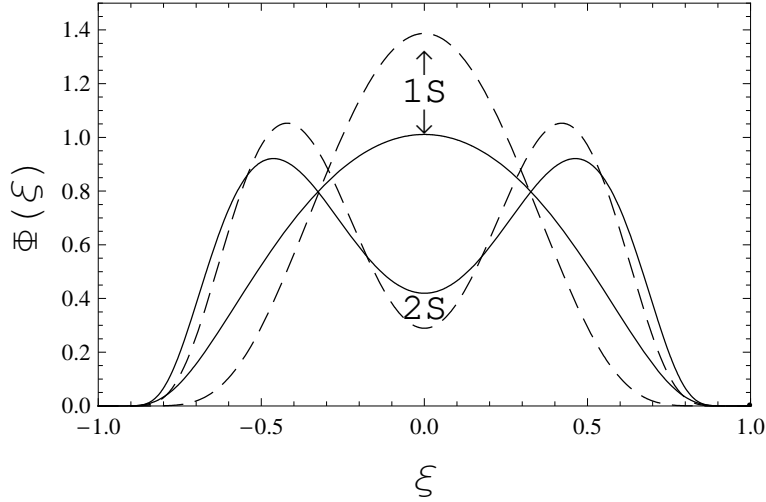}
\caption{Quark distribution amplitudes $\Phi_{\eta_c}(\xi)$ and
$\Phi_{\eta'_c}(\xi)$ of this work (solid lines) and that of Ref.
\cite{Braguta1,Braguta2} (dashed lines).}
 \label{fig:xi}
\end{figure}
For the latter, the equation \cite{Braguta1,Braguta2}
 \be \label{eq:Braguta}
 \Phi(\xi,\mu=1.2\text{GeV})=c(\alpha,\beta,\gamma)(1-\xi^2)(\alpha+\gamma\xi^2)~{\rm
 exp}\left(-\frac{\beta}{1-\xi^2}\right),
 \en
was used to plot the curves. $\Phi^{1S}(\xi,\mu)$ and
$\Phi^{2S}(\xi,\mu)$ corresponded to $(\alpha=1,\beta=3.8,\gamma=0)$
\cite{Braguta1} and $(\alpha=0.027,\beta=2.49,\gamma=1)$
\cite{Braguta2}, respectively. In Fig. 3, the major difference
between the two curves of the $1S$ states was that the curve of Ref.
\cite{Braguta1}, on which $\xi$ is peaked around zero, was sharper
than that of this work. This meant that the momentum fraction $x$ in
the function used in Ref. \cite{Braguta1} was more centered on $1/2$
than in the Gaussian-type wave function. As to the curves of the
$2S$ state, we found the locations of extreme value by
differentiating both Eqs. (\ref{LC2S}) and (\ref{eq:Braguta}) over
$\xi$. The results were
 \be
 \xi&=& 0, \pm \sqrt{\frac{3}{3+4 d}},\non \\
 \xi&=& 0, \pm \frac{1}{2 \sqrt{\gamma}}[3\gamma +\beta\gamma-\alpha-
 (\alpha^2+2 \alpha\gamma+6
 \alpha\beta\gamma+\gamma^2+6\beta\gamma^2+\beta^2\gamma^2)^{1/2}]^{1/2},\non
 \en
respectively. The values are $\xi=0, \pm 0.463$ for this work and
$\xi=0, \pm 0.421$ for Ref. \cite{Braguta2}. For the bottom sector,
the quark distribution amplitudes $\Phi^{(_{'})1S}_{P,V}(\xi)$ and
$\Phi^{(_{'})2S}_{P,V}(\xi)$ are plotted in Figs. 4 and 5.
\begin{figure}
\includegraphics*[width=4in]{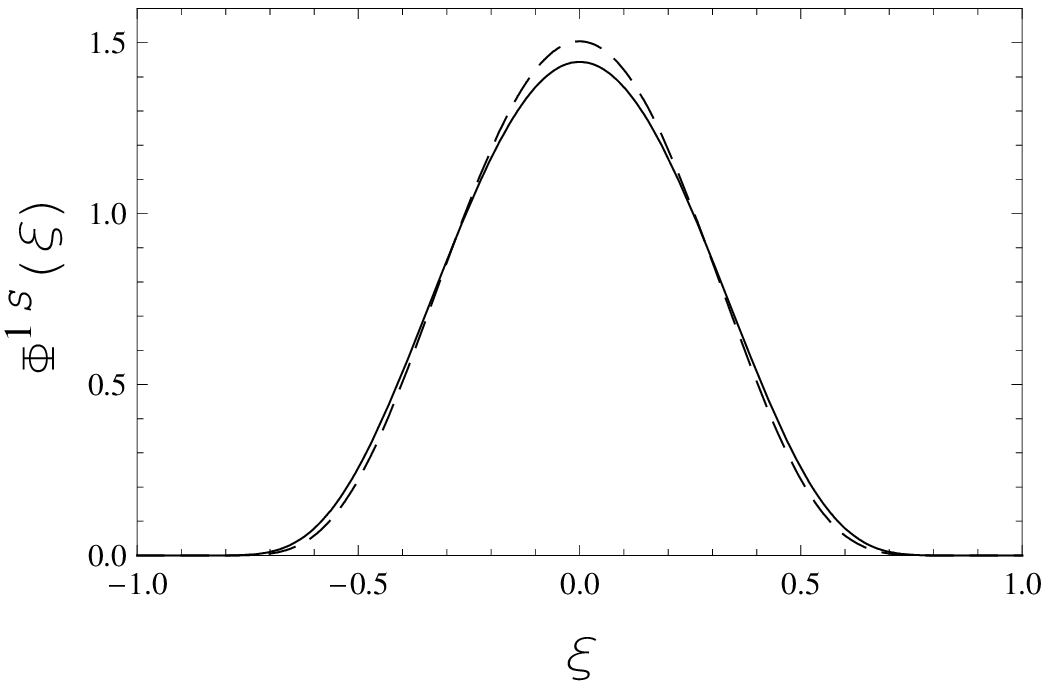}
\caption{ Quark distribution amplitudes $\Phi_{\eta_b}(\xi)$ (solid
lines) and $\Phi'_{\Upsilon}(\xi)$ (dashed lines) of this work.}
 \label{fig:xib1S}
\end{figure}
\begin{figure}
\includegraphics*[width=4in]{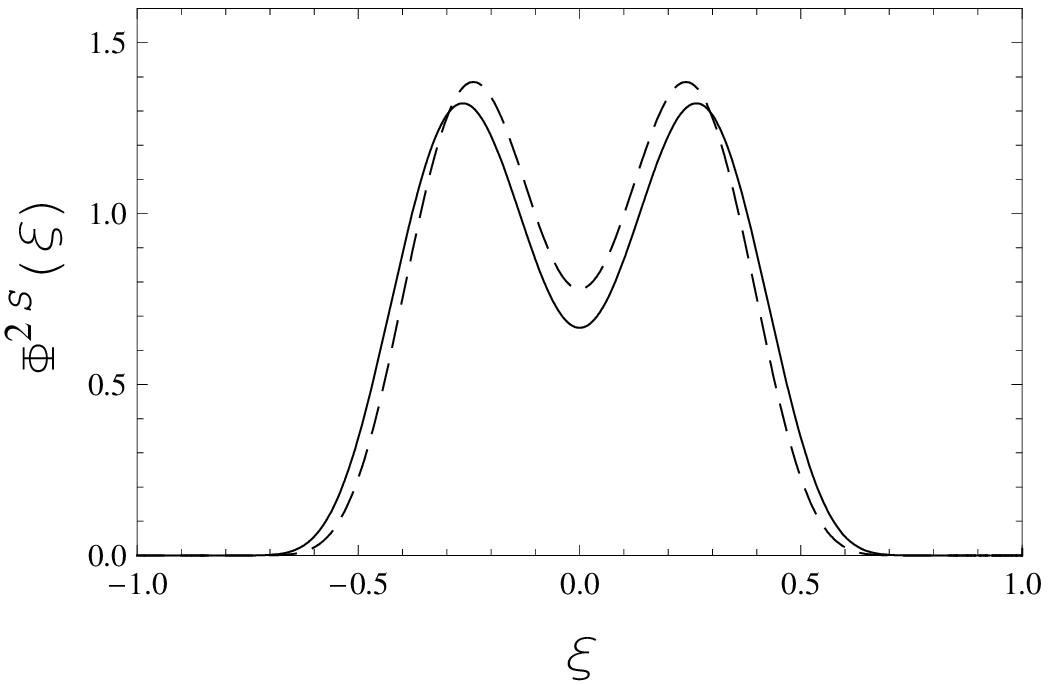}
\caption{ Quark distribution amplitudes $\Phi_{\eta'_b}(\xi)$ (solid
lines) and $\Phi'_{\Upsilon'}(\xi)$ (dashed lines) of this work.}
 \label{fig:xib2S}
\end{figure}
These figures show that the $x$-distribution of the pseudoscalar
bottomonium was almost the same as for the vector bottomonium. The
reason is that since the differences between the pseudoscalar and
the vector heavy quarkonium states arise from $1/m_Q$ corrections,
the larger the $m_Q$, the smaller the difference between them. For
comparison, the $x$-distributions of the charmonium and bottomonium
states are plotted together in Fig. 6.
\begin{figure}
\includegraphics*[width=4in]{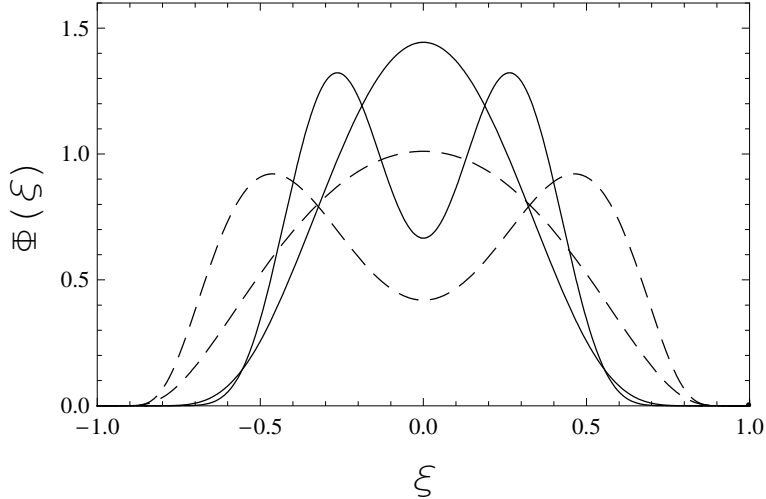}
\caption{ Quark distribution amplitudes $\Phi_{\eta_b,\eta'_b}(\xi)$
(solid lines) and $\Phi_{\eta_c,\eta'_c}(\xi)$ (dashed lines) of
this work.}
 \label{fig:xibc1s2S}
\end{figure}

In addition, the differences between various quark distribution
amplitudes can also be shown by $\xi$-moments. We calculated the
$\xi$-moments Eqs. (\ref{ximoment1s}), (\ref{ximoment2s}),
(\ref{ximomentV1s}), and (\ref{ximomentV2s}). Our results and those
of other theoretical groups are listed in Tables III and IV. In
these tables, Refs. \cite{Braguta1,Braguta2,Braguta3} used the QCD
sum rules. The authors of Ref. \cite{BT} calculated in the framework
of the Buchmuller-Tye-potential model and found that their results
are in agreement with experiments, which included the leptonic
widths and hyperfine splittings. The authors of \cite{EGKLY} made
calculations for the Cornell potential. The predictions of
light-cone sum rules are shown in Ref. \cite{BC}. The authors of
Ref. \cite{CJ} not only numerically calculated the $1S$ charmonium
states in conventional LFQM, but they also used the variational
approach to fix the mass of the charm quark. The authors of Ref.
\cite{BF} constructed quark distribution amplitudes by assuming the
very non-relativistic limit, where $\Phi(x) = \delta(x-1/2)$, and
then redistributing the parton momenta by relativistic gluon
exchange. Depending on the assumed value of $\alpha_s$ at the heavy
quark scale, they found the $\xi^n$ moments of $\eta_c$ in Table
III. For reasonable values of $\alpha_s$ they obtained qualitatively
similar values as in \cite{Braguta1} and the curves of them are
slightly
narrower than in our numerical analysis. 
As to the $\xi$-moments for the bottomonium states, they were
evaluated and are listed in Table V.
\begin{table}[h!]
\caption{\label{tab:1sxi} The $\xi$-moments for the $1S$ charmonium
states. ($^\dag$ $\langle \xi^n \rangle_{\eta_c}=\langle \xi^n
\rangle_{J/\psi}$) }
\begin{ruledtabular}
\begin{tabular}{c|ll|ll|ll}
 moment & $\langle \xi^2 \rangle_{\eta_c}$ & $\langle \xi^2 \rangle_{J/\psi}$ & $\langle \xi^4 \rangle_{\eta_c}$
 & $\langle \xi^4 \rangle_{J/\psi}$ & $\langle \xi^6 \rangle_{\eta_c}$ & $\langle \xi^6 \rangle_{J/\psi}$
  \\ \hline
 this work & $0.117^{+0.010}_{-0.011}$  & $0.0966^{+0.0013}_{-0.0013}$ & $0.0307^{+0.0052}_{-0.0049}$
 & $0.0215^{+0.0005}_{-0.0005}$ & $0.0109^{+0.0026}_{-0.0023}$ & $0.00657^{+0.00023}_{-0.00023}$ \\
  \cite{Braguta1,Braguta3}& $0.070^{+0.007}_{-0.007}$ & $0.072^{+0.007}_{-0.007}$
  & $0.012^{+0.002}_{-0.002}$ & $0.012^{+0.002}_{-0.002}$ & $0.0032^{+0.0009}_{-0.0009}$ & $0.0033^{+0.0007}_{-0.0007}$\\
  \cite{BT} & & $0.086$ & &$0.020$ & & $0.0066$\\
  \cite{EGKLY}& & $0.084$ & & $0.019$ && $0.0066$ \\
  \cite{BC}$^\dag$ & & $0.13$ &  &$0.040$ & &  $0.018$ \\
  \cite{CJ} & $0.084^{+0.004}_{-0.007}$& $0.082^{+0.004}_{-0.006}$& $0.017^{+0.001}_{-0.003}$& $0.016^{+0.002}_{-0.002}$
  &$0.0047^{+0.0006}_{-0.0010}$&$0.0046^{+0.0005}_{-0.0010}$\\
  \cite{BF} & $0.067$ & & $0.011$& &$0.004$ &
\end{tabular}
\end{ruledtabular}
\end{table}
\begin{table}[h!]
\caption{\label{tab:2sxi} The $\xi$-moments for the $2S$ charmonium
states. ($^\dag$ $\langle \xi^n \rangle_{\eta'_c}=\langle \xi^n
\rangle_{\psi'}$)}
\begin{ruledtabular}
\begin{tabular}{c|ll|ll|ll}
 moment & $\langle \xi^2 \rangle_{\eta'_c}$ & $\langle \xi^2 \rangle_{\psi'}$ & $\langle \xi^4 \rangle_{\eta'_c}$
 & $\langle \xi^4 \rangle_{\psi'}$ & $\langle \xi^6 \rangle_{\eta'_c}$ & $\langle \xi^6 \rangle_{\psi'}$
  \\ \hline
 this work & $0.192^{+0.008}_{-0.008}$ & $0.136^{+0.006}_{-0.006}$& $0.0600^{+0.0048}_{-0.0044}$ &
 $0.0320^{+0.0021}_{-0.0020}$ & $0.0229^{+0.0026}_{-0.0023}$ & $0.00950^{+0.00078}_{-0.00074}$ \\
 \cite{Braguta2}$^{\dag}$&  & $0.18^{+0.05}_{-0.07}$ &  & $0.051^{+0.031}_{-0.031}$& &
 $0.017^{+0.016}_{-0.014}$\\
  \cite{BT}&   &$0.16$  & & $0.042$ & & $0.015$\\
\end{tabular}
\end{ruledtabular}
\end{table}
\begin{table}[h!]
\caption{\label{tab:1s2sxib} The $\xi$-moments for the $1S$ and $2S$
bottomonium states. }
\begin{ruledtabular}
\begin{tabular}{c|ll|ll|ll}
 moment & $\langle \xi^2 \rangle_{\eta_b}$ & $\langle \xi^2 \rangle_{\Upsilon}$ & $\langle \xi^4 \rangle_{\eta_b}$
 & $\langle \xi^4 \rangle_{\Upsilon}$ & $\langle \xi^6 \rangle_{\eta_b}$ & $\langle \xi^6 \rangle_{\Upsilon}$
  \\ \hline
 this work & $0.0643^{+0.0005}_{-0.0005}$  & $0.0598^{+0.0004}_{-0.0004}$ & $0.0103^{+0.0001}_{-0.0001}$
 & $0.00894^{+0.00011}_{-0.00011}$ & $0.00237^{+0.00005}_{-0.00005}$ & $0.00192^{+0.00003}_{-0.00003}$ \\ \hline\hline
 moment & $\langle \xi^2 \rangle_{\eta'_b}$ & $\langle \xi^2 \rangle_{\Upsilon'}$ & $\langle \xi^4 \rangle_{\eta'_b}$
 & $\langle \xi^4 \rangle_{\Upsilon'}$ & $\langle \xi^6 \rangle_{\eta'_b}$ & $\langle \xi^6 \rangle_{\Upsilon'}$
  \\ \hline
 this work & $0.0844^{+0.0010}_{-0.0010}$  & $0.0729^{+0.0007}_{-0.0007}$ & $0.0128^{+0.0003}_{-0.0003}$
 & $0.00984^{+0.00018}_{-0.00018}$ & $0.00255^{+0.00009}_{-0.00008}$ & $0.00176^{+0.00005}_{-0.00004}$ \\
\end{tabular}
\end{ruledtabular}
\end{table}

\section{Conclusions}
This study performed the transverse momenta $p_\perp$ integrals of
formulae for the decay constants $f_{P,V}$ of $1S$ and $2S$ heavy
quarkonium states and then obtained their quark distribution
amplitudes $\Phi_{P,V}(\xi)$. In addition, the $\xi$-moments
$\langle \xi^{2,4,6} \rangle_{P,V}$ were also obtained by
integrating out $\xi$ in $\Phi_{P,V}(\xi)$. For each heavy
quarkonium state, the five parameters $m_Q$, $\beta_P$,
$\beta_{P'}$, $\beta_{V}$, $\beta_{V'}$ which appeared in the
momentum distribution amplitude were determined. This study first
extracted the decay constants $f_{V,V'}$ from the experimental data
of the leptonic decay $Br(V\to e^+e^-)$, and it then used the Van
Royen-Weisskopf formula to obtain the decay constants $f_{P,P'}$.
These decay constants were used as constraints to fix the above
parameters. Then, the curves of the quark distribution amplitudes
for the $1S$ and $2S$ heavy quarkonium states were plotted by the
fixed parameters. It was found that, for the $1S$ charmonium state,
the momentum fraction $x$ in the function used by Ref.
\cite{Braguta1} was more centered on $1/2$ than the one in the
Gaussian-type wave function. In addition, the $x$-distribution of
the pseudoscalar bottomonium was almost the same as that of the
vector bottomonium. The reason for this was that the differences
between these heavy quarkonium states, which arise from $1/m_Q$
corrections, become small when $m_Q$ is large. Finally, the
numerical results of the $\xi$-moments were calculated and compared
with the experimental data and other theoretical predictions. 

{\bf Acknowledgements}\\
The author would like to thank Hsiang-nan Li, Tsung-Wen Yeh, and
Alexey Luchinsky for their helpful discussions.
 This work is supported in part by the National Science Council of R.O.C. under Grant No
 NSC-96-2112-M-017-002-MY3.



\begin{thebibliography}{99}

 \bi{PDG08} C. Amsler {\it et al.} (Particle Data Group), Phys. Lett. {\bf B 667}, 1 (2008).

 \bibitem{EGMR} E. Eichten {\it et al.}, FERMILAB-PUB-07-006-T, [hep-ph/0701208v3].

 \bi{KEDR} V. M. Aulchenko {\it et al.} (KEDR), Phys. Lett. {\bf B
 573}, 63 (2003).

 \bi{CLEO} D. M. Asner {\it et al.} (CLEO), \prl {\bf 92}, 142001
 (2004).

 \bi{BABAR} B. Anbert {\it et al.} (BABAR), \prl {\bf 92}, 142002 (2004).

 \bi{BABAR1} B. Anbert {\it et al.} (BABAR), \prl {\bf 101}, 071801
 (2008).

 \bibitem{QR} C. Quigg and J. L. Rosner, Phys. Rept. {\bf 56}, 167-235
 (1979).

 \bibitem{BPP} S. J. Brodsky, H. C. Pauli and S. S. Pinsky, Phys. Rept. {\bf 301}, 299 (1998).

 \bibitem{LFQM} M. V. Terent'ev, Sov. J. Phys. {\bf 24}, 106 (1976).

 \bi{LFQMa} V. B. Berestetsky and M. V. Terent'ev, Sov. J. Phys. {\bf 24}, 547
 (1976).

 \bi{LFQMb} M. V. Terent'ev, Sov. J. Phys. {\bf 25}, 347 (1977).

 \bibitem{Jaus1} W. Jaus, Phys. Rev. D {\bf 41}, 3394 (1990).

 \bi{Jausa} W. Jaus, \prd {\bf 44}, 2851 (1991).

 \bibitem{CCH1} H. Y. Cheng, C. Y. Cheung and C. W. Hwang, Phys. Rev. D {\bf 55}, 1159 (1997).

 \bibitem{Jaus2} W. Jaus, \prd {\bf 60}, 054026 (1999).

 \bibitem{CCH2} H. Y. Cheng, C. K. Chua and C. W. Hwang, Phys. Rev. D {\bf 69}, 074025 (2004).

 \bi{Hwang} C. W. Hwang, \prd {\bf 64}, 034011 (2001).

 \bi{hwangwei} C. W. Hwang and Z. T. Wei, J. Phys. G {\bf 34}, 687
 (2007).

 \bi{Braguta1} V. V. Braguta, A. K. Likhoded, and A. V. Luchinsky, Phys. Lett. B {\bf 646}, 80
 (2007).

 \bi{Braguta2} V. V. Braguta, \prd {\bf 77}, 034026 (2008).

 \bi{Braguta3} V. V. Braguta, \prd {\bf 75}, 094016 (2007).

 \bi{BT} W. Buchmuller and S. H. H. Tye, \prd {\bf 24}, 132 (1981).

 \bi{EGKLY} E. Eichten, K. Gottfried, T. Kinoshita, K. D. Lane, and T. -M. Yan,
 \prd {\bf 17}, 3090 (1978).

 \bi{BC} A. E. Bondar and V. L. Chernyak, Phys. Lett. B {\bf 612}, 215
 (2005).


 \bi{CJ} H. M. Choi and C. R. Ji, \prd {\bf 76}, 094010 (2007).

 \bi{BF} G. Bell and T. Feldmann, JHEP {\bf 0804}, 061 (2008).

 \bibitem{CM69} S. J. Chang and S. K. Ma,
 Phys.\ Rev.\  {\bf 180}, 1506 (1969).

 \bi{ISGW} N. Isgur, D. Scora, B. Grinstein, and M. B. Wise, \prd
 {\bf 39}, 799 (1989).


 \bi{CZ} V. L. Chernyak and A. R. Zhitnitsky, Phys. Rep. {\bf 112},
 173 (1984).

 \bi{BHL} S. J. Brodsky, T. Huang, and G. P. Lepage, in Banff 1981,
 Proceedings, Particles and Fields 2, pp. 143.


 \bi{DM} D. Ebert and A. P. Martynenko, \prd {\bf 74}, 054008
 (2006).



 \bi{NS} M. Neubert and B. Stech, Adv. Ser. Direct. High Energy
 Phys. {\bf 15}, 294-344 (1998).

 \bi{DT}  N. G. Deshpande and J. Trampetic, Phys. Lett. B {\bf 339}, 270
 (1994).

 \bi{VRW} R. Van Royen and V. F. Weisskopf, Nuovo Cimento {\bf 50}, 617
 (1967).


 \bi{newdata} R. E. Mitchell {\it et al.} (CLEO), Phys. Rev. Lett. {\bf 102}, 011801 (2009) .

\end{thebibliography}
\end{document}